\documentstyle[12pt,epsf]{article}

\newcommand{\ds}{\displaystyle}
\newcommand{\ra}{\rightarrow}
\newcommand{\be}{\begin{equation}}
\newcommand{\ee}{\end{equation}}
\newcommand{\bea}{\begin{eqnarray}}
\newcommand{\eea}{\end{eqnarray}}
\newcommand{\ci}{\cite}
\newcommand{\bi}{\bibitem}
\newcommand{\nono}{\nonumber \\}

\newcommand{\s}{\sigma}

\def\dal{\,\lower0.3ex\vbox{\hrule\hbox{\vrule\kern2pt\vbox{\kern4pt\kern4pt}
\kern2pt\vrule}\hrule}\,}

\def\s{\sigma}

\begin{document}

\title{\sl Wavepacket diffraction in the Kronig-Penney model}
\vspace{1 true cm}
\author{G. K\"albermann$^*$
\\Soil and Water dept., Faculty of
Agriculture, Rehovot 76100, Israel}
\maketitle

\begin{abstract}

The phenomenon of wavepacket diffraction in space and time is
investigated numerically and analytically, 
for a one-dimensional array of equally spaced finite-depth wells.
Theoretical predictions for the lattice at 
long times and at low scattering energies,
coincide exactly with the results for a single well.
At intermediate and short times compared to the classical passage time,
the pattern shows both a broad diffractive pattern and an interference
pattern inside each diffractive peak.
The diffractive structure persists for this case to infinite time.
\end{abstract}


$^*${\sl e-mail address: hope@vms.huji.ac.il}

\newpage

\section{\sl Introduction}

In previous publications \ci{k1,k2,k3} the 
phenomenon of wavepacket diffraction in space and time was
described, both numerically in one and two dimensions \ci{k1,k2},
 as well as analytically for one and three dimensions at long times.\ci{k3}
The phenomenon was found in wavepacket potential scattering for the
 nonrelativistic, Schr\"odinger equation and for the relativistic Dirac
equation.

The essential characteristic of the phenomenon consists in the production of
diffractive peak structure that travels in space, due to the
interference between the incoming and spreading wavepacket and the scattered 
 wave.
The patterns are produced by a time independent potential.

The peak structure exists for all packets, but, it survives 
only for packets that are initially narrower than a value related to the
potential width (see below). For wider packets, the peak structure
merges into a single peak.

The effect is named  {\sl wavepacket diffraction in space and time}.

For plane wave monochromatic waves, quantum scattering displays
diffraction phenomena in time \ci{mosh} induced by the sudden
opening of a slit, or in space by fixed slits or gratings.
The combined effect of time dependent opening of slits
for plane monochromatic waves produces diffraction patterns in space and
time.\ci{zeilinger}
These patterns fade out asymptotically in time.
Recent measurements of atomic wave diffraction\ci{prl},
 have indeed demonstrated that the {\sl diffraction in time},
process is supported by experiments.

It was found in \ci{k1,k2,k3} that there is no need to
 have a sudden (time-dependent) opening of a slit to generate
a time-dependent diffractive train. The multiple peak traveling structure
 is generated by a time independent potential provided the packet
 width is narrower than the extent of the potential, either well or
barrier. The time dependent behavior stems from the time-dependent
spreading of the initial and scattered packets.

The scattering events investigated so far considered only a single well.
However, any target, be it a foil that is irradiated by some beam, or
 a quantum well or heterostructure scattering a current, involves a
large array of wells.
The purpose of the present work is to address the problem of wavepacket
 diffraction in the context of a one-dimensional array of equally
 spaced finite size and fine depth wells, as the
 time-honored Kronig-Penney model\ci{kittel}. Our interest will
be in dealing with a finite size array and not an infinite one.
 Therefore Bloch's theorem will not be of a help in treating the
scattering event theoretically. The phenomenon we treat depends on
the spatial extent of the wells. We will therefore avoid using $\delta$
function type of potentials. It will be seen that, despite these
 complications, the outcome of the scattering is quite independent of
the number of wells, their, depth, etc. The only aspect that will
matter crucially is the width of a single well in the array. 
The results open up the possibility of using the phenomenon in solid
 state devices.
In the next section we summarize some results of ref.\ci{k3}. Section 3 
deals with the Kronig-Penney model analytically, 
and section 4 numerically. Section 5 brings a brief summary.

\section{\sl Wavepacket diffraction in space and time}

In\ci{k3} we found that, for $t\ra\infty$ it is possible to
 write analytical formulae for the
diffraction pattern produced by a scattering event of a wave packet at
 a well or barrier. 
The formulae were derived 
for the simplest possible case that can be dealt with almost completely
analytically. This is the example of a wave packet scattering off
a square well in one and three dimensions.
The results, however, were shown to be independent of the shape
of the well and the packet.

As it is our intention to treat the one dimensional array of wells, 
we recapitulate here the one dimensional formula as a bridge to 
 the lattice case.

Consider a gaussian wave packet impinging from the left,
on a well located around the origin,

\bea\label{sqwell}
V(x)= -V_0~\Theta(w/2-|x|)
\eea

where, {\sl w} is the width, $\Theta$ is the Heaviside function, 
and $V_0$, the depth.
Using the results of \ci{bohm}, 
the reflected and transmitted packets become

\bea\label{psifull}
\psi(x,t)=~\int_{-\infty}^{\infty}\phi(k,x,t)~a(k,q_0)~
e^{-i~\frac{k^2}{2~m}~t}dk
\eea
where $\phi(x,t,q_0)$ is the stationary solution to the square well scattering
problem for each $\sl k$ and $\sl a(k,q_0)$ is the Fourier transform
amplitude for the initial wave function with average momentum $\sl q_0$.

As the effect is due to the interference between the
 incoming spread packet and the reflected wave we focus on the
the backward direction for $\sl x<-w/2$

\bea\label{psi}
\phi(k,x,t)&=&D(k,k')~e^{i~k~(x-x_0)}+~F(k,k')~e^{-i~k~(x+x_0+2~w)}\nono
D(k,k')&=&1\nono
F(k,k')&=&\frac{E(k,k')}{A(k,k')}\nono
E(k,k')&=&-2~i~(k^2-k'^2)~sin(2~k'w)\nono
A(k,k')&=&(k+k')^2~e^{-2~i~k'~w}-(k-k')^2~e^{2~i~k'~w}
\eea

where 
\bea\label{form}
a(k,q_0)=e^{i~k~(x-x_0)-\s^2~(k-q_0)^2}
\eea
with $\ds k'=\sqrt{k^2+2~m~|V_0|}$, and $\s$, the width parameter
of the packet.

At $k\approx 0$ we have $F\approx-1$.
For long times and distances $x>>x_0>>w$ we find using the properties of 
Gaussian integrals

\bea\label{asymp}
|\psi|&=&2~\sqrt{\frac{2~m~\pi}{t}}~e^{-z}~|sin\bigg(\frac{m~x}{t}(x_0
+2~i~\s^2~q_0)\bigg)|\nono
&=&2~\sqrt{\frac{2~m~\pi}{t}}~e^{-z}~\sqrt{sin^2(\frac{m~x~x_0}{t})+
sinh^2(\frac{2~\s^2~q_0~m~x}{t})}\nono
z&=&\s^2~\bigg(\frac{m^2~(x^2+x_0^2)}{t^2}+q_0^2\bigg)
\eea

This expression represents a diffraction pattern that travels in time
and persists.
In \ci{k3}, we compared the expression above to the numerical solution.
Excellent agrrement was found without resorting to 
any scale factor adjustment.(See figure 1 in \ci{k3}).

We further derived the condition for the pattern to persist to be

\bea\label{constraint}
\s<<\sqrt{\frac{w}{q_0}}
\eea

A relation between the initial width of the packet, the width of the
 well or barrier and the initial average momentum of the packet.
Narrow packets only display the effect.

We connected the results to the process of diffraction in time\ci{mosh},
 and noted that the phenomenon of diffraction in space and time with
wave packets did not need a time dependent opening of a slit, a
 time independent potential sufficed. It was found that the
 pattern did not fade out at long times as in the case of the 
diffraction in time\ci{mosh}, and it applied to any potential, be
it repulsive or attractive.

The key element for the obtention of the formula of eq.(\ref{asymp}), 
was the replacement of
{\sl F} in eq.(\ref{psi}), by the value at threshold, namely $F=-1$. This
replacement is valid for long times at which the wild oscillations
in the exponentials demand a very low momentum. 
Due to the fact that this value is independent of the type of well
 or barrier, the result holds in general for any kind of well or barrier.
In all cases where the reflection coefficient {\sl F} can be replaced 
by the value at threshold we obtain a receding multiple peak coherent
 wave train or, a single hump, depending on the initial width of the packet.
This is the essence of the phenomenon of {\sl wavepacket diffraction in space
and time}.

\section{\sl Kronig-Penney scattering}

Consider now, a one dimensional array of {\sl N} square wells of width {\sl a}
 to the right of the origin, 
a distance $\delta+a$ apart from each other. We shift the
location of the wells as compared to eq.(\ref{sqwell}) for convenience.

\bea\label{sqwells}
V(x)= -V_0\sum_{n=0}^{N-1}~\Theta(\frac{a}{2}-|x-l_n-\frac{a}{2}|)
\eea

where the left edge of the well is $l_n=n~(\delta+a)$, 
while the right edge is located at $r_n~=~l_n+a$.

In order to show that, this one dimensional Kronig-Penney lattice of
finite size and depth wells yields essentially
 the same result as a single well, 
we will expand the reflection coefficient for the lattice
to third order in {\sl k}. 

In order to find the reflection coefficient, equivalent
to {\sl F} of eq.(\ref{psi}) we have to solve for the wave function as
a function of {\sl k} by means of plane waves, without resorting
to Bloch's theorem that applies only for the infinite array.

For each well we write

\bea\label{psiwells}
\psi&=&A_n~e^{i~k~x}+~B_n~e^{-i~k~x}~~~~~~~~~~~~r_{n-1}<x<l_n\nono
&=&C_n~e^{i~\kappa~x}+~D_n~e^{-i~\kappa~x}~~~~~~~~~~~~l_n<x<r_n\nono
&=&A_{n+1}~e^{i~k~x}+~B_{n+1}~e^{-i~k~x}~~~~~~~~~~~~r_n<x<l_{n+1}
\eea

with $\kappa=\sqrt{k^2+2~m~V_0}$, {\sl m} being the mass of
the impinging particle. For the last well we demand, an outgoing 
wave, hence $B_{N}=0$.

Continuity of the wave function and its derivarive 
at the right and left edges of each well gives

\bea\label{transfer}
\left(\begin{array}{c}A_n \\ B_n\end{array} \right)&=&M_n
\left(\begin{array}{c}A_{n+1} \\ B_{n+1}\end{array} \right)\nono
M_n&=&\left(\begin{array}{cc}
z_1~e^{i~k~a}&-z_2~e^{-i~k~(r_n+l_n)}\\
z_2~e^{i~k~(r_n+l_n)}&z_1^*~e^{-i~k~a}\end{array}\right)
\eea

where $z_1=cos(\kappa~a)-i~sin(\kappa~a)~(1+u^2)/(2~u)$, 
$z_2=i~sin(\kappa~a)~(1-u^2)/(2~u)$, $u=\frac{k}{\kappa}$.

Due to the strong oscillations of $e^{-i~\frac{k^2}{2~m}~t}$
in eq.(\ref{psifull}), the long time behavior of the wave function, 
is determined by the very low $u=\frac{k}{\kappa}$ regime.
We therefore expand the matrices in powers of {\sl u}.
The reflextion coefficient {\sl F}, will be then obtained
 from $ F=\frac{B_0}{A_0}$ as a polynomial in {\sl u}.

For a plane wave stationary solution with no imcoming waves from the
 rightmost edge of the wells and the number of wells $N>>2$, 
we find after some lengthy algebra

\bea\label{ab}
\frac{A_0}{A_n}&=&1+c_0\lambda+(N-2)~c_0\lambda(1-e^{4~i~k\delta})
+O(\epsilon^2)\nono
\frac{B_0}{A_n}&=&-(1+2~c_0\lambda+(N-2)~c_0\lambda(1-e^{4~i~k\delta})
+O(\epsilon^2)
\eea

where $c_0=cos(\kappa_0~a)$, $\ds \lambda=\frac{\epsilon}{-i~sin(\kappa_0~a)~
(1-e^{2~i~k~\delta})}$, $\epsilon=\frac{k}{\kappa_0}$, and
 $\kappa_0=\sqrt{2~m~V_0}$ is real for wells, 
and imaginary for barriers.
From the expressions above it is seen that to this order, the correction to
$F\approx -1$ originates from the term multiplied
by (N-2). However this term is identical for both the expressions above.
Therefore, to this order, $A_0$ and $ -B_0$ are almost identical.

We have carried out an expansion to $O(\epsilon^3)$ also. The expression
 becomes quite involved having some 20 terms each for $A_0$ and $B_0$. 
Even to this order, the leading terms still give $F\approx -1$.

The Kronig-Penney lattice displays the same low 
{\sl k} behavior of
the reflection coefficient as a single well. 

In light of the results of the previous section,
 it follows that the Kronig-Penney lattice will generate scattered wave packets 
with the same characteristics as a single well. In particular, 
the phenomenon of 
diffraction of wavepackets in space and time \ci{k3} will apply 
to the lattice also.
In the next section we will show numerical calculations that support this 
claim, and also depict scattering events
 for different number of wells at finite times in contrast to the
results found here for long times.

\section{\sl Numerical results for the Kronig-Penney lattice}

The following series of pictures depict the amplitude
 of the wave function obtained with the numerical algorithm described
 in \ci{k1}. We use Gaussian wells defined by

\bea\label{gauswell}
V(x)=-V_0~\sum_{n=0}^{N-1}exp\bigg({-\frac{(x-x_n)^2}{b^2}}\bigg)
\eea

Where the equivalence to the square wells array of eq.(\ref{sqwells})
is given by $b\approx\frac{w}{2}$.
The impinging packet is a gaussian wave packet 
traveling from the left
with an average speed $v$, initial location $x_0$, mass {\sl m}, wave number
$q~=~m~v$ and width $\s$,

\bea\label{packet}
\psi~=~C~exp\bigg({i~q~(x-x_0)-\frac{(x-x_0)^2}{4~\s^2}}\bigg)
\eea

We choose the parameters
 to be $V_0=1$ $a=1$, $x_n = 8$, $m=20$, $q=1$, $\s=0.5$, and $x_0=-20$, 
that implement the constraint of eq.(\ref{constraint}).
 In the case of a wider packet we find a smooth hump proceeding 
backwards and forwards \ci{k3}.

Figures 1,2,3 show, the backwards scattered, the forward 
scattered and the well region wave amplitudes at t=15000 for various numbers of 
wells.
\begin{figure}
\epsffile{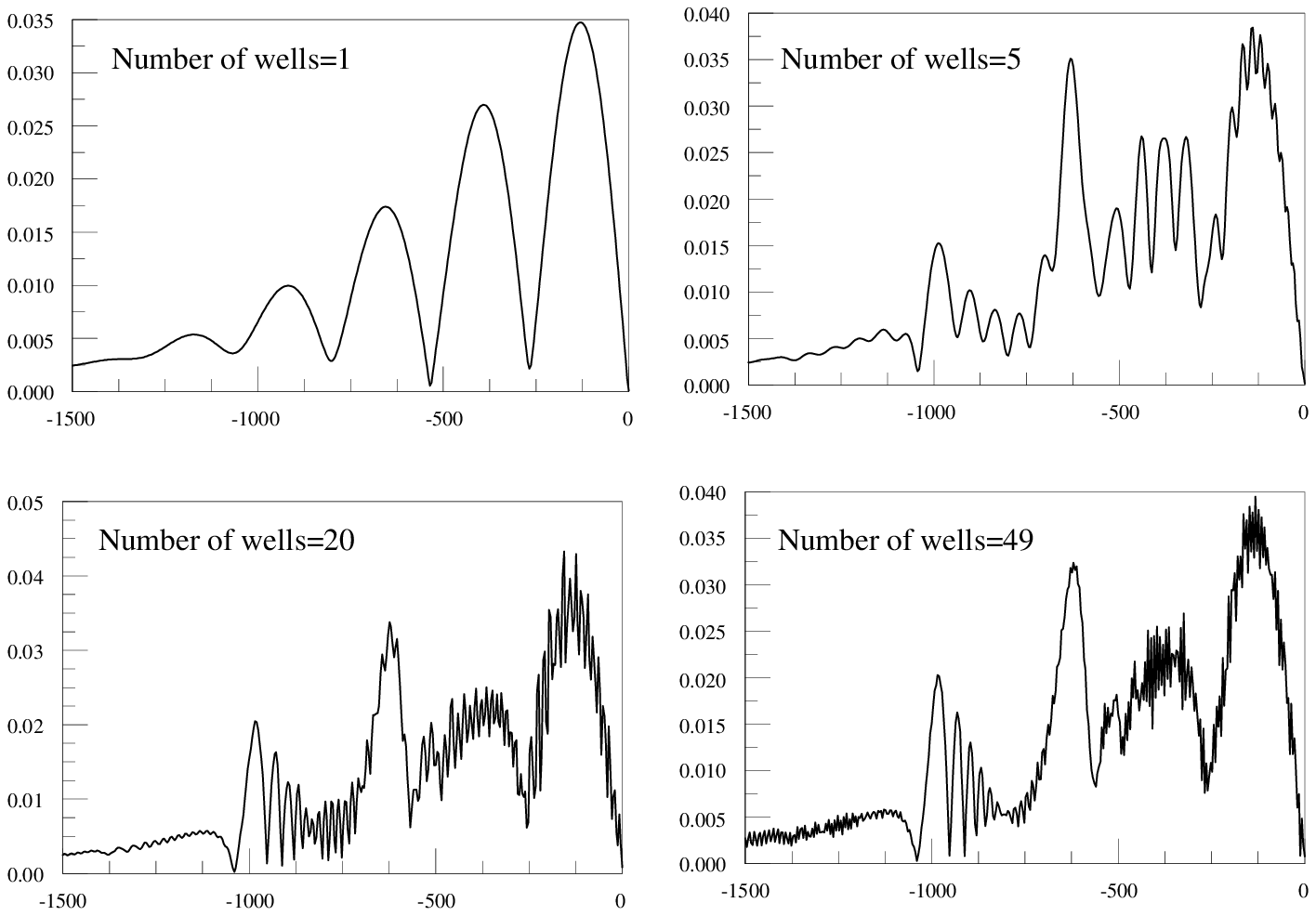}
\vsize=5 cm
\caption{\sl Wave amplitude in the backward region for different numbers
of wells.}
\label{fig1}
\end{figure}
\begin{figure}
\epsffile{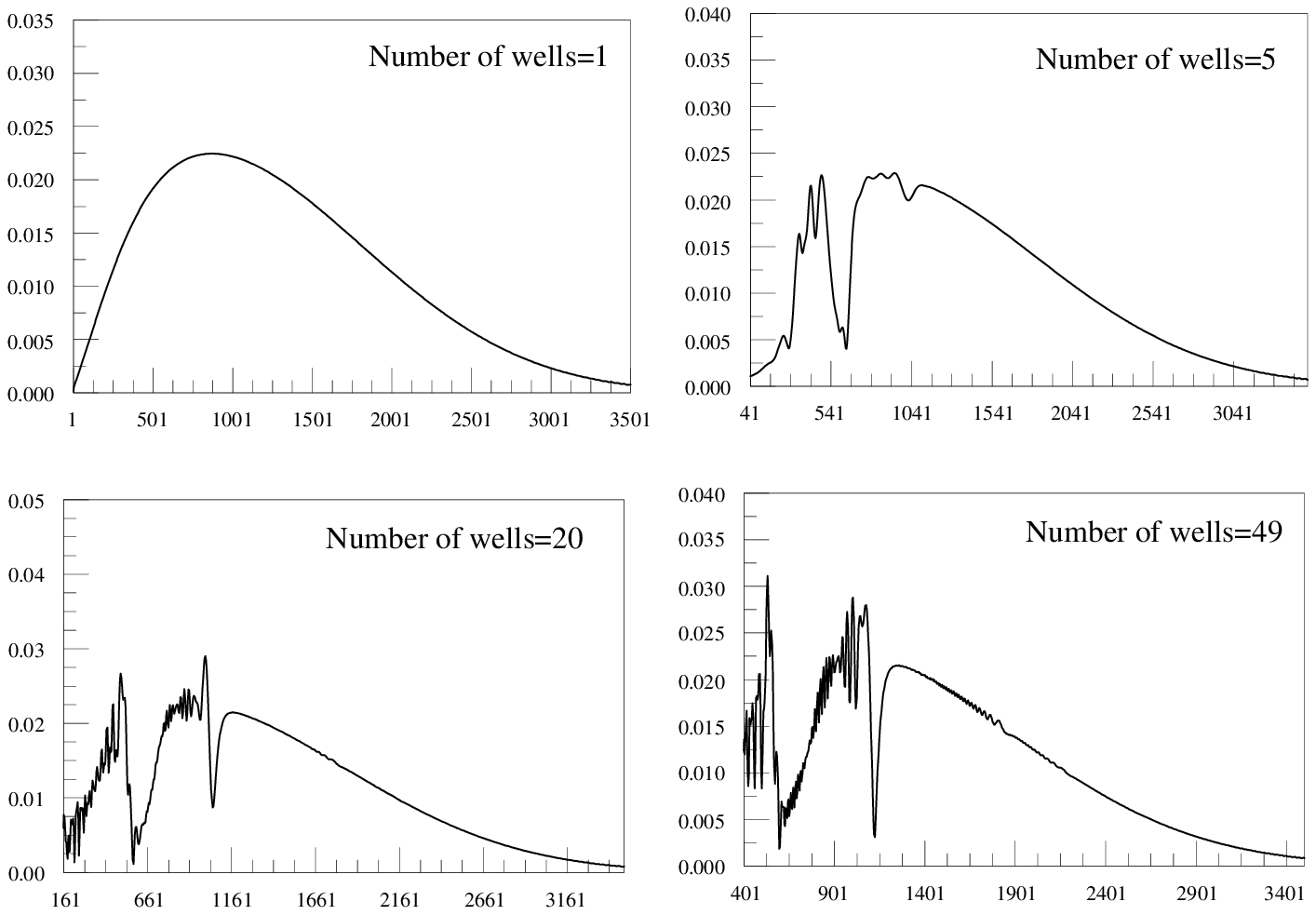}
\vsize=5 cm
\caption{\sl Wave amplitude in the forward region for different numbers
of wells.}
\label{fig2}
\end{figure}
\begin{figure}
\epsffile{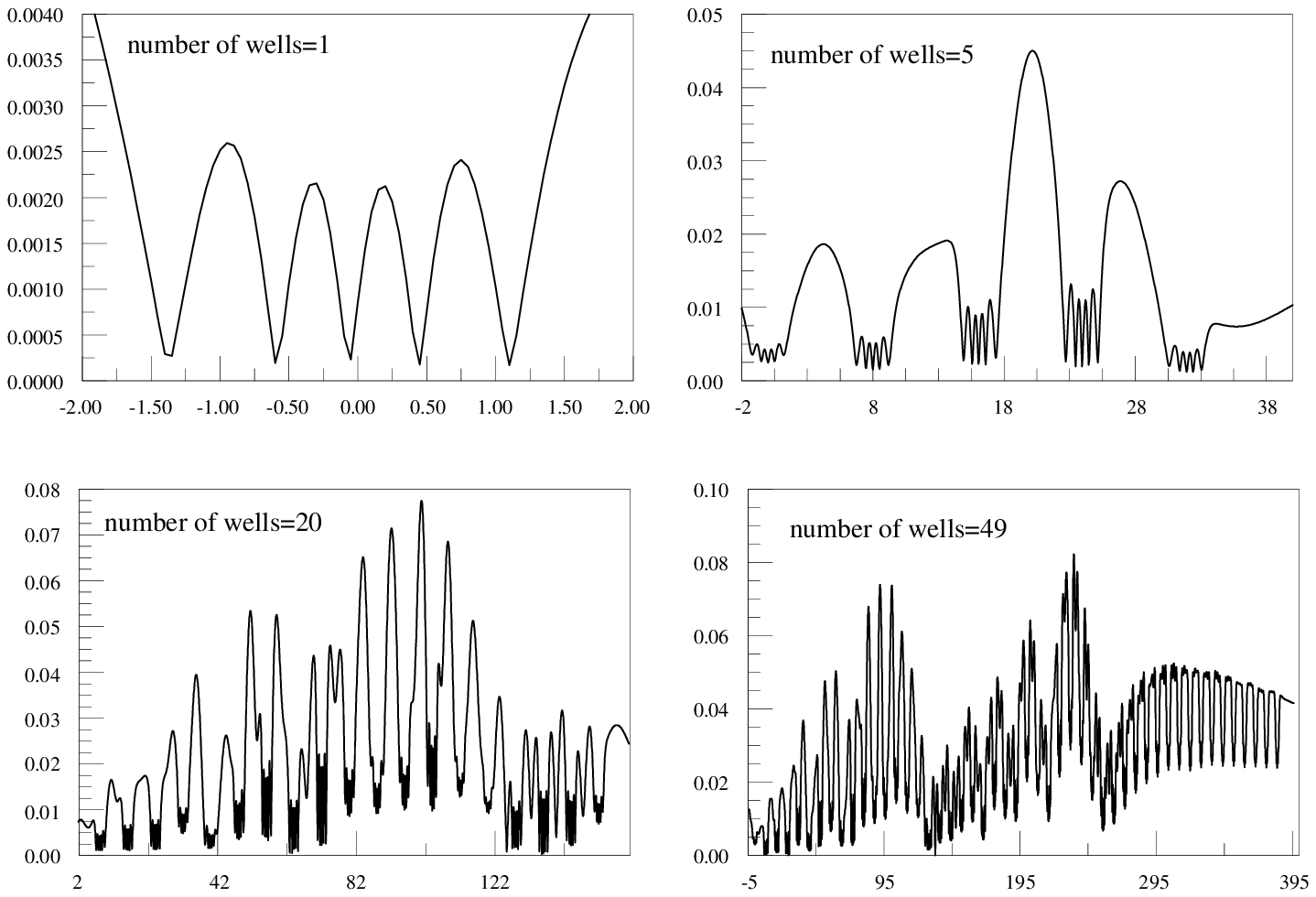}
\vsize=5 cm
\caption{\sl Wave amplitude in the wells region for different numbers
of wells.}
\label{fig3}
\end{figure}

The figures show clearly that the diffraction pattern that dominates the
 wave profile in the backwards direction is the one of a single well, as
 expected from the results of the previous section.

The amplitude of the wave is modulated by the diffraction pattern,  
with the interference pattern appearing as oscillations inside each broad
 peak. The number of oscillations is
identical to the number of wells, higher order interferences
 are blurred out into the background.

In figure 4 we depict a detail of the wave inside the 12$^th$ well for
the case of 49 wells as compared to the wave for the case of
a  single scatterer.
The behavior is almost identical. The pattern repeats itself inside
 each well of the lattice, but the amplitude varies in space (well number),
 and time.

\begin{figure}
\epsffile{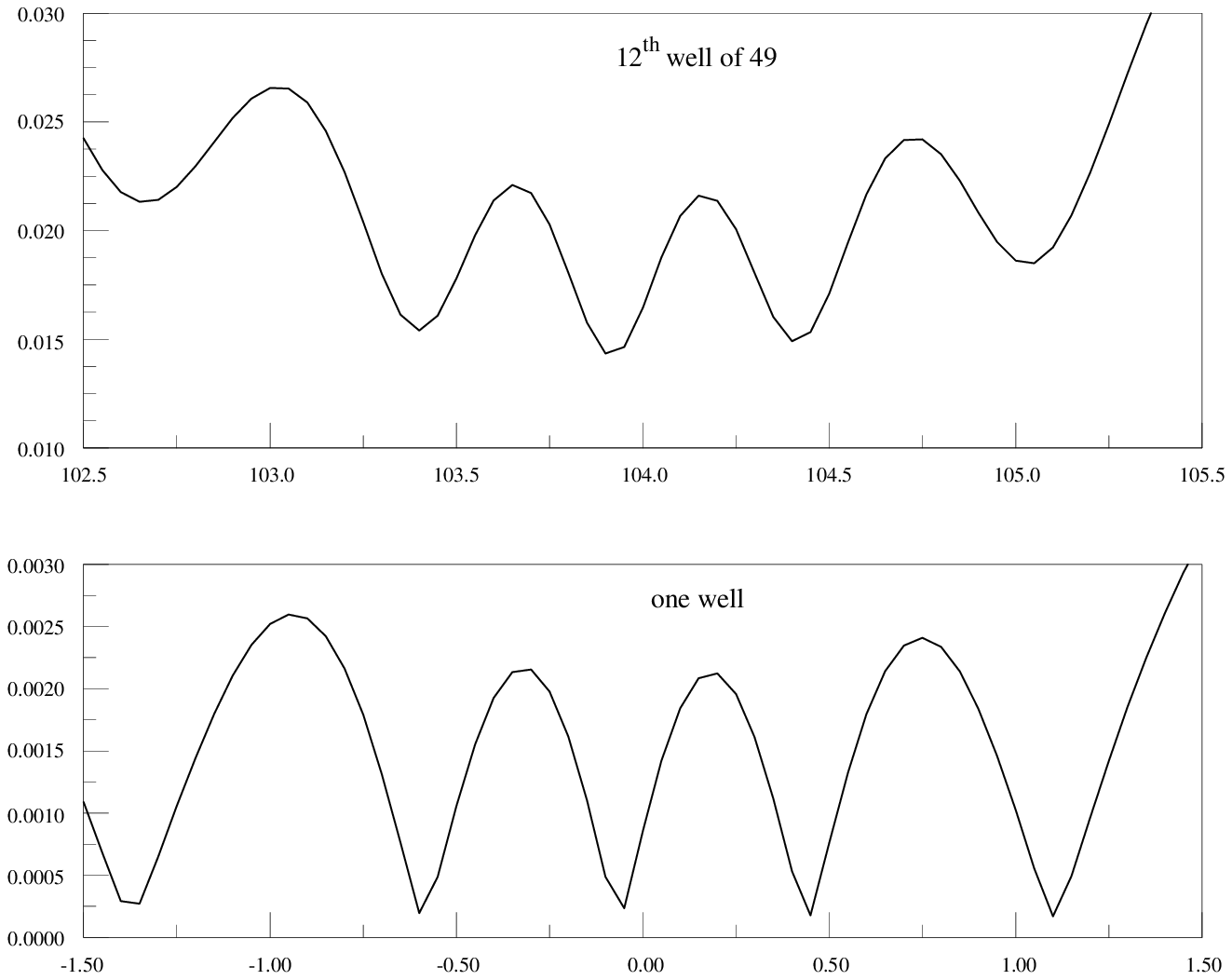}
\vsize=5 cm
\caption{\sl Detail of the wave amplitude inside a certain well.}
\label{fig4}
\end{figure}

In the previous section, we found that the long time behavior
 should be dominated by the diffraction pattern of a single well.
Therefore, the interference ripples should diminish as time elapses.

Figure 5 shows the results for a longer time and 49 wells.
This calculation demanded almost a day of computer work. Further
increase in the time makes the calculation prohibitive.
However, this increase should be enough in order to see if 
there is a trend of dissipation of the interference ripples.

Indeed, the dissapearance of the
broken peaks is most evident at long distance, at intermediate
 distances it is hard to assess if this is the case. We can expect that at
 even longer times the broken peaks will converge into the diffraction
 pattern humps broadening them to the size obtained with a single well.

\begin{figure}
\epsffile{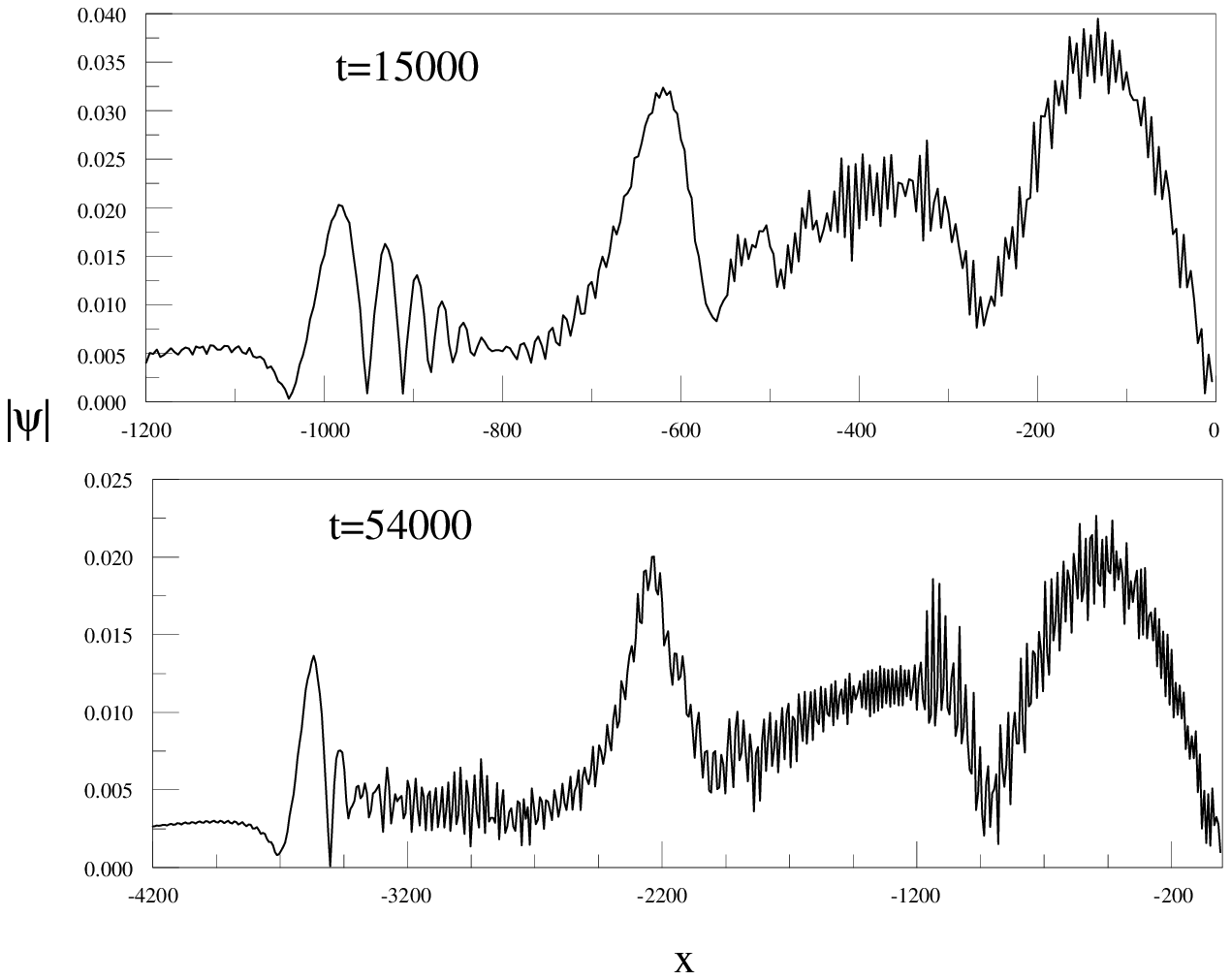}
\vsize=5 cm
\caption{\sl Wave amplitude in the backward region for 49 wells at different
times.}
\label{fig5}
\end{figure}

\section{\sl Brief summary}

We have found that the phenomenon of wavepacket 
diffraction in space and time remains intact for a one
dimensional array of wells. 
On the experimental side, the present results suggest that besides the
experiment simulated in \ci{k3} with Helium, it could be possible
to use solid state devices, such as an array of quantum wells to 
implement the effect.
Future work will focus on the more realistic situation of a higher dimensional
lattice of scatterers.

\end{document}